\documentclass[prb,aps,twocolumn,draft,tightenlines,%
showpacs,floatfix]{revtex4}  

\usepackage{graphicx}
\usepackage{dcolumn}
\usepackage{bm}
\usepackage{amssymb}
\usepackage{amsmath}
\usepackage{psfig}
\begin{document}

\title{Formation and decay of electron-hole droplets in diamond}
\author{J. H. Jiang}
\author{M. W. Wu}%
\thanks{Author to whom correspondence should be addressed}%
\email{mwwu@ustc.edu.cn.}%
\affiliation{Hefei National Laboratory for Physical Sciences at
  Microscale, University of Science and
  Technology of China,  Hefei, Anhui, 230026, China and\\
Department of Physics, University of Science and
Technology of China, Hefei, Anhui, 230026, China}
\altaffiliation{Mailing Address.}
\author{M. Nagai}
\affiliation{Department of Physics, Kyoto University, Kyoto 606-8502, Japan}
\author{M. Kuwata-Gonokami}
\affiliation{Department of Applied Physics, University of Tokyo, Hongo, Bunkyo-ku, Tokyo 113-8656, Japan}

\date{\today}

\begin{abstract}
  We study the formation and decay of electron-hole droplets in 
  diamond at  low and high temperatures under different excitations by 
  master equations. The calculation
  reveals that at low temperature the kinetics of the system
 is similar to that of a direct
  semiconductor, whereas 
  at high temperature it is metastable 
and similar to an indirect semiconductor. 
  Our results at low temperature are consistent with the
  experimental findings reported
  by Nagai {\em et al.} [Phys. Rev. B {\bf 68}, 081202 (R) (2003)].
  The kinetics of the e-h system in diamonds at high temperature under
  both low and high excitations is also predicted.
\end{abstract}
\maketitle

Photoexcited electron-hole (e-h) systems in semiconductors provide a
unique opportunity to understand quantum many-body phenomena with
Coulomb interactions. In the dilute density region, an electron and a hole are
combined to form a neutral bound state---an exciton. At low
temperature, a dense exciton gas condenses into a liquid
 phase  with a
metallic character in the form of e-h droplets (EHDs). 
This macroscopic metallic phase has been extensively investigated in
the past three decades in indirect
semiconductors such as Ge and Si.\cite{Keldysh} 
The transition between EHD and exciton gas
is considered analogous to a classical liquid
transition in  water, and the EHD formation is 
well understood with a
classical nucleation theory.\cite{WestBook} However, the formation
and decay of EHD in photoexcited semiconductors are not only
determined by the collection and evaporation of excitons on the surface of
EHD but also by carrier recombination. 
In direct semiconductors in particular, a fast recombination process overcomes
the thermal kinetics of carriers. Therefore e-h pairs 
annihilate before a small
e-h ensemble grows to become a macroscopic-size
EHD. Consequently, the phase
transition is shown to be of second order.\cite{2nd_order1}

This competition between thermal kinetics and recombination of
carriers is also apparent in indirect semiconductors. 
In traditional indirect semiconductors such as Ge and 
Si, at a certain high temperature the evaporation rate is larger than 
the recombination rate, which makes the kinetics of EHD formation 
similar to that of classical nucleation. In this case, EHD
formation exhibits a hysteresis effect and the average drop size is
large. However when the temperature is sufficiently low, the thermal
kinetics is suppressed. The dominant recombination effect
makes the e-h system behave like those in direct
semiconductors, {\em i.e.}, 
no hysteresis effect and a small average number of 
pairs per cluster (ANPC).
Under this condition the exciton-EHD phase transition
changes from first- to
second-order.\cite{2nd_order1,2nd_order2}
This density and temperature
region, where the thermo-dynamical phase diagram is distorted, is
attractive to scientists because the quantum statistics of the
quasi-particles is dominant, and a hidden collective phase including
Bose-Einstein condensation might appear. In order to understand such a rich
variety of macroscopic phases, it is important first to
evaluate the kinetics of the liquid-gas transition in the photoexcited
indirect semiconductors in the low temperature region. 
Nevertheless it is not very realistic for conventional 
indirect semiconductors because of their narrowness in energy scale.

Diamond is a wide band gap indirect semiconductor with a band
structure similar to those of Ge and Si, and is a good
candidate to study 
carrier dynamics. Moreover, because of the
small dielectric constant of
diamond the screening of the Coulomb interaction
between carriers is small.
Thus one can treat e-h system in diamond in  wide energy scale. 
Recently Shimano {\em et al.} evaluated the character of EHD
in diamond by 
time-resolved luminescence measurements and reported a higher critical
temperature, larger work function, larger density, and shorter
lifetime for EHD in diamond compared to Ge and Si.\cite{High Tc} 
Consistent values are also obtained from an analysis of the luminescence
spectra under quasi-cw excitation.\cite{Thonke1,Thonke2} 
The dynamics of the EHD formation at 12 K under a different excitation
density has also been studied by Nagai {\em et
  al}.\cite{PRB} It was observed that after photoexcited 
carriers are cooled rapidly into
a supersaturated exciton gas within several tens of picosecond, spatial
condensation of dense exciton gas into EHD occurs within a few hundred
picoseconds.
In this report we investigate theoretically the
kinetics of EHD formation and decay in diamond.
First we use a discrete master equation theory developed by  Haug and 
Abraham\cite{Haug4 direct-gap} to investigate the  femtosecond 
excitation in diamond at a low temperature regime where only 
small e-h  clusters are formed. We then use the continuous  
 master equation theory developed by Silver\cite{Silver} and
by Koch and Haug\cite{Haug3 indirect-gap1} to investigate the
dynamics
at high temperature regime where the average drop size is too large to
be treated discretely. Finally we compare our results with the
experimental measurements by Nagai {\em et al.}\cite{PRB}
The division between the low and high temperature regimes
in diamond  is $\sim$60\ K
where the  thermal evaporation rate equals to the recombination 
rate.\cite{60K}

For a discrete master equation formalism, if the 
concentration of clusters containing $n$ e-h pairs at time $t$ 
is denoted by $f(n,t)$,  the master equation
describing the evolution of $f(n,t)$ is:
\begin{equation}
  \frac{\partial}{\partial t}f(n,t) = j_{n-1} - j_{n}  
\label{jn}
\end{equation}
for $n\ge 2$, where $j_n$ is the net probability current between the
clusters with $n$ and $n+1$ e-h pairs:
\begin{equation}
  j_n = g_n f(n,t) - l_{n+1} f(n+1,t)\ .
\label{gn}
\end{equation}
In this equation $l_n$ and $g_n$ are the gain and loss rates of a cluster 
with $n$ e-h pairs. The gain rate is obtained from the assumption 
that excitons with a density $n_x$ are collected at the surface 
of a cluster, and is approximated by $g_n = b n_x n^{2/3}$
with $b = 4 \pi R_0^2 v_x$. $R_0=(3/4\pi \rho_0)^{1/3}$ is the 
Wigner-Seitz radius of the EHD and 
$v_x=\sqrt{{k T}/{2 \pi m_x}}$ is the thermal velocity of 
excitons with an effective mass $m_x$.  $\rho_0$ denotes the EHD density.
The loss rate is composed of the sum of the evaporation rate $\alpha _n $ and 
the recombination rate $n/\tau _n$.  
$l_n = \alpha _n + n/\tau _n$.
The evaporation rate is given by a time-independent Richardson-Dushman current, 
$\alpha _n = b D_x \exp[(-\phi + c\sigma n^{-1/3})/(kT)] n^{2/3}$,
where  $D_x =  \gamma_x ({m_x k T}/
{2 \pi \hbar^2})^{3/2}$ is the effective 
density of state of exciton, 
$\gamma_x$ is the degeneracy of the exciton ground state and
$c\sigma n^{-1/3}$ represents the correction of the 
binding energy due to surface
effect with $\sigma$ denoting the surface energy of the EHD. 
These equations are solved together with the continuity equation:
\begin{equation}
  \frac{\partial}{\partial t}f(1,t) =
  G(t)-\sum_{n=1}^\infty\frac{nf(n,t)}{\tau_n} -2j_1-\sum_{n=2}^\infty
  j_n\ ,
\end{equation}
with $G(t)$ representing the excitation pulse.
\begin{equation}
  G(t)=G_0 e^{-(t-t_0)^2/t_p^2}
\end{equation}
with $t_p$ standing for the width of laser pulse which is 0.1\
ps throughout this paper.
It is noted that the coalescence of clusters larger than excitons is
neglected.
\widetext

\begin{table}[bth]
  \centering
  \caption{EHD and exciton parameters for diamond which
    are used in the calculation}
  \begin{tabular}{l|c|c|c|c} \hline
    & \ symbol \ & value & \ \ \ unit \ \ \ & Ref. \\ \hline
    mean EHD lifetime 
    & $\tau_d$ & 1 & ns & \onlinecite{High Tc}\\
    mean exciton lifetime & $\tau_x$ & 100 & ns &  \\
    work function of EHD & $\phi$ & 50 & meV & \onlinecite{High Tc,Thonke2,parameter} \\
    surface energy of EHD & $\sigma_0$ & 1.2 & erg/cm$^2$ & \onlinecite{sigma}\\
    $\sigma(T)=\sigma_0(1-(T/T_c)^2)$ & $T_c$ & 165 & K & \onlinecite{High Tc}\\
    exciton degeneracy & $\gamma$ & 12 & & \\
    effective mass of exciton & $ m_x $ & 7.92 & 10$^{-31}$ kg & \\
    e-h density of EHD & $\rho_0$ & 1.0 & 10$^{20}$ cm$^{-3}$ &
    \onlinecite{High Tc}\\
    \hline
  \end{tabular}
  \label{table2}
\end{table}
\endwidetext

For the case of high temperature where $n$ is too large to be treated
discretely, we turn to the equation of moments.\cite{Haug3 indirect-gap1}
The $\nu$-th moment of EHD distribution is defined 
as
\begin{equation}
\label{moment}
  x_{\nu}(t) = \int\limits^{\infty}_{n_c} n^{\nu} f(n,t)\,dn
\end{equation}
in which $n_c$ is a critical size where the stationary
distribution has a minimum. All clusters smaller than $n_c$ are
counted as excitons, while clusters larger than $n_c$ are treated as
EHDs. Under this approximation, the ``exciton density'' is given by:
\begin{equation}
\label{nxt}
n_x(t) =\int\limits^{n_c}_{1} n f(n,t)\,dn\ .
\end{equation}
Here $n_c$ is calculated approximately by  equating loss rate and the
gain rate with the recombination loss neglected:
\begin{equation}
\alpha _{n_c} \approx n_x(t) b n_c^{2/3}.
\label{n_c}
\end{equation}
The equations of moments are given by:
\begin{eqnarray}
  \frac{d}{dt} x_0 &=& J_{n_c} - (\frac{d}
  {dt}n_c) f(n_c,t)\ ,
  \label{12}\\
  \frac{d}{dt} x_{\nu}& =& n_c^{\nu} \frac{d}
  {dt}x_0 - \nu [ \frac{x_{\nu}}{\tau _d} + x_{\nu-1/3} b 
  ( n_s - n_x ) ]\ ,
\end{eqnarray}
with $\tau _d$ denoting the mean EHD lifetime and $n_s$ being the
saturated exciton density, $n_s = D_x \exp ( -\phi /kT)$.
$\nu = 1/3$, 2/3, 1, 4/3, 5/3, and 2.
Finally the continuity equation is
\begin{equation}
\label{cont}
  \frac{d}{d t} n_x = - \frac{n_x}{\tau _x} - n_c
  \frac{d}{d t} x_0 + x_{2/3} b (n_s-n_x) + G(t)\ ,
\end{equation}
in which $\tau _x$  represents the mean exciton lifetime.
The second term describes the change in EHD density and the
third term is due to free exciton evaporation and collection by EHD.
Equations\ (\ref{12}) to (\ref{cont}) 
form a closed set of equations. The expressions
of $J_{n_c}$ and $f(n_c,t)$ in Eq.\ (\ref{12}) are given
in Appendix A.

We first study the kinetics of the e-h system in diamond at  
12\ K by using the  discrete master equations, under the
experimental conditions similar to the ones used 
by Nagai {\em et al}.\cite{PRB} 
The material parameters are listed in TABLE I.\cite{exciton}
The results of our calculation  are plotted in Figs.\ 1-4.

\begin{figure}
  \centerline
  {\psfig{figure=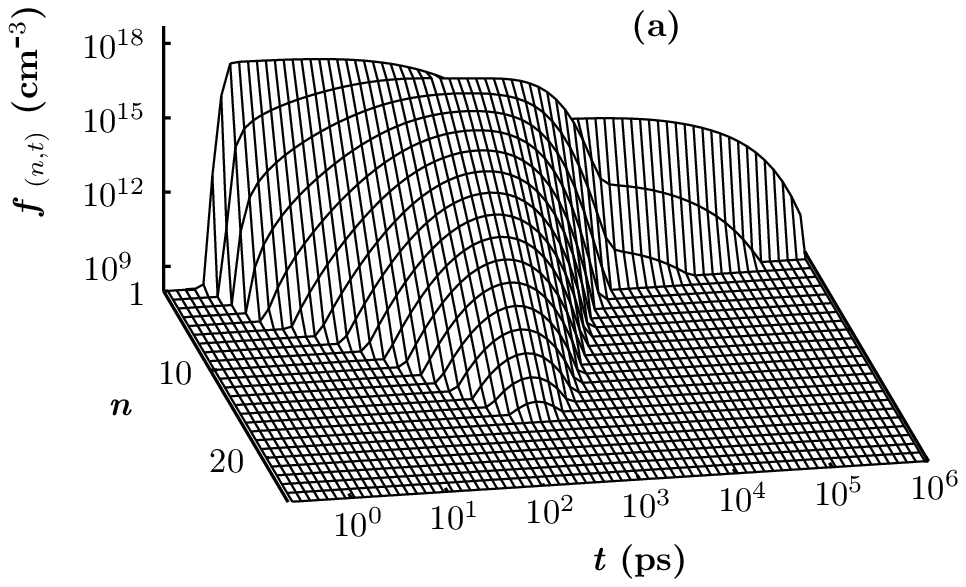,width=8cm,height=4.8cm,angle=0}}
  \bigskip
  \centerline
  {\psfig{figure=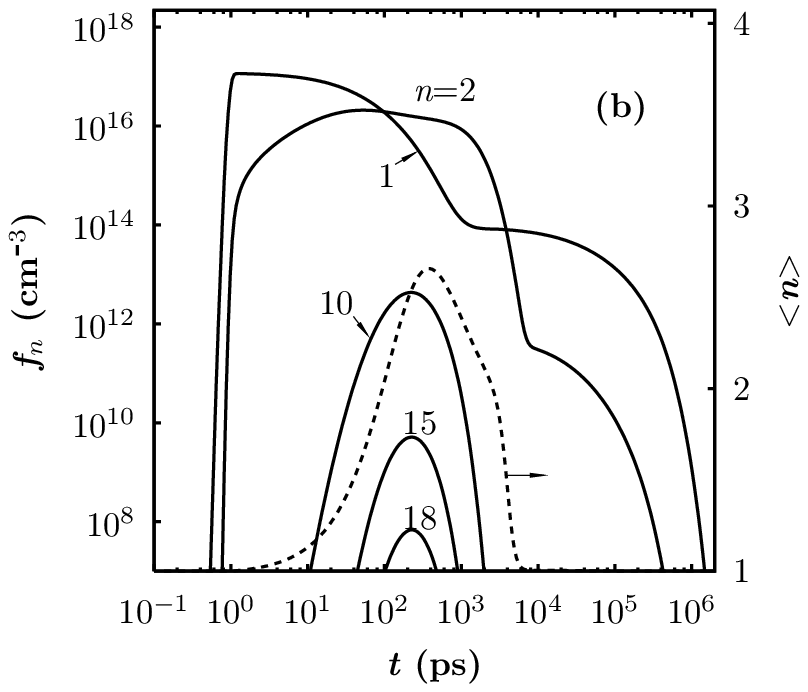,width=7.2cm,height=6cm,angle=0}}
  \caption{(a): Cluster concentration versus time and number of e-h 
      pairs per cluster  for diamond at $T=12$ K under Gaussian pulse 
    excitation with $G_0 = 6.6\times 10^{29}$
    cm$^{-3}$s$^{-1}$,$t_p = 0.1$ ps, and $t_0 = 1.0$ ps. 
    (b): Concentration of some selected
      clusters  (solid curve)
    and ANPC $\langle n\rangle$ (dashed curve) versus time.
    }
  \label{fig1}
\end{figure}

In Fig.\ 1 we present the kinetics under an
excitation with $G_0 = 6.6\times 10^{29}$\ cm$^{-3}$s$^{-1}$
and $t_p=0.1$\ ps which corresponds to a
$\sim$0.2\ mJ/cm$^2$ excitation in the 
experiment.\cite{PRB,exci} 
The cluster concentration versus time and the number
of e-h pairs in a cluster as well as the concentration of some
selected clusters  versus time
are plotted in Figs.\ 1(a) and (b) 
respectively. The figures indicate that the concentration of
small clusters rise faster than that of large cluster because 
  the current $j_n$ 
in Eq.\  (\ref{jn}) flows
from small clusters to larger ones. It is also seen from the figures
that  the system reaches
quasi-equilibrium at about 120\ ps and this
quasi-equilibrium lasts 
about 200\ ps. 

In the experiment the peak energy of the EHD
emission band shifts toward the low energy side during first 200 ps,\cite{PRB}
which suggests that large clusters
are formed at  a longer time. Meanwhile from the
fact that there is little change of the
luminescence in 
the low energy regime, we conclude that the rate of 
the formation of small clusters from excitons and 
the rate of coalescence to large clusters are nearly the same.
As a result  the concentration of small clusters rises and quickly reaches a 
steady value. These features are consistent with our calculation.
The luminescence indicates that the system reaches its 
quasi-equilibrium in about 200\ ps, which is
comparable to the 120 ps value we obtained. 
It is also seen in Fig.\ 1(b) that excitons ($n=1$) decay much 
slower than clusters with $n \ge 2$---because the lifetime of excitons $\tau
_x$ is larger than that of e-h pairs in e-h clusters $\tau _d$  
and also because excitons get additional compensation from the recombination 
of large clusters.

In order to monitor the average size of the cluster, we  plot 
the ANPC  $\langle n\rangle = 
{\sum^{\infty}_{n=1} n f(n,t)/\sum^{\infty}_{n=1}f(n,t)}$ 
as a function of time in Fig.\ 1(b).  The ANPC is less
than 3, and is different
to that obtained for direct band-gap  semiconductors.\cite{Jiang,Haug4 direct-gap}
Note that in the present model the coalescence of  clusters larger than
excitons ({\em e.g.} biexcitons) is neglected. The
coalescence of
clusters adds to the cluster formation mechanism, thus this approximation
leads to a smaller ANPC in our results.
Despite this, the ANPC is still too small to result in the
  formation of macroscopic EHDs.
The smallness of ANPC and the shortness of the
time during which the system is in quasi-equilibrium indicate 
that the system is characteristic of a nonequilibrium, 
similar to an e-h system in direct semiconductors. 

\begin{figure}[htb]
  \centerline{\psfig{figure=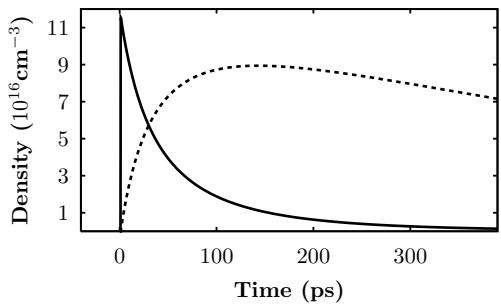,width=6.6cm,height=4.cm,angle=0}}
  \caption{Exciton density $n_x$ (solid curve)  and the total density of
    e-h pairs in all clusters larger than exciton $n_d$ (dashed curve) 
    for diamond at $T=12$ K under the same excitation as in FIG. 1.}
  \label{fig2}
\end{figure}

We also show  the time evolution of the exciton density
$n_x$ and the total density of all e-h pairs condensed in 
clusters larger than excitons: 
$n_d = \sum^{\infty}_{n=2} n f(n,t)$
in Fig.\ 2. Thus we can 
compare the evolution of these densities with that 
of excitons and the integrated EHD luminescence intensities. We can 
see  that excitons slowly condense into EHDs and 
$n_d$ reaches a maximum  around 150\ ps, which
corresponds to the 260\ ps experimental data. 

There are some differences between these densities and the 
luminescence intensities in the experiment.
First, the times when exciton and EHD densities reach their maxima 
are about one half  those in the experiment. 
This  originates partly from the simplified excitation model we use. 
In this model the only excitations generated by the laser pulse 
are excitons. In reality, these excitations
should be e-h plasma and they are
always overheated.  After the excitation the e-h system is cooled down
in several tens of picoseconds.\cite{PRB}
This relaxation process affects the kinetics of
exciton. The  absence of this process results in a shorter
formation time in the calculation. 
Second, at the equilibrium stage 90\% of the excitation is converted into
the EHD phase, while in the experiment  only about
 50\%  is converted.
This may be partly due to the diffusion of the
  e-h pairs,
and therefore the actual exciton density is smaller than our
evaluation. Nevertheless 
the diffusion effect might be marginal 
since the time scale is only 300 ps and  the initial carrier 
distribution with a penetration  depth of 15 $\mu$m  is less spatially
inhomogeneous.
It is reported that
the effect of diffusion of an e-h system in Si is
negligible  during the 200 ps after
excitation.\cite{JPSJ} Moreover, the high density of e-h pairs causes large Auger 
recombination in EHD and 
a repulsion of excitons from EHD by the phonon wind.\cite{phonon} 
Thus the efficiency of the collection of excitons which collide with
  EHD decreases,
and less excitons are converted into EHD. 
These effects also slow down the formation
process as the gain rate $g_n$ in Eq.\ (\ref{gn}) is proportional to
exciton density, and thus lead to a longer formation time in the
experiment compared to our results.

\begin{figure}[htb]
  \centerline
  {\psfig{figure=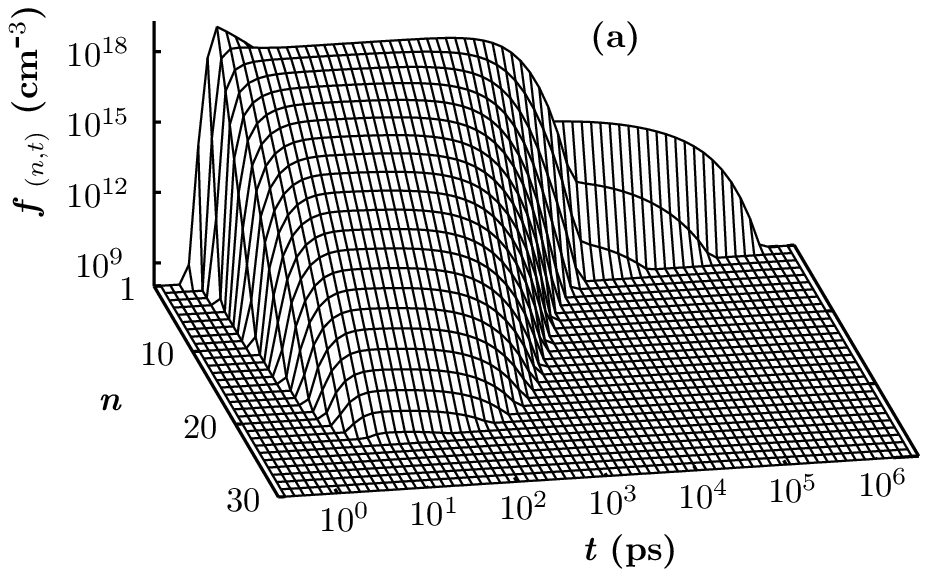,width=8cm,height=4.8cm,angle=0}}
  \bigskip
  \centerline
  {\psfig{figure=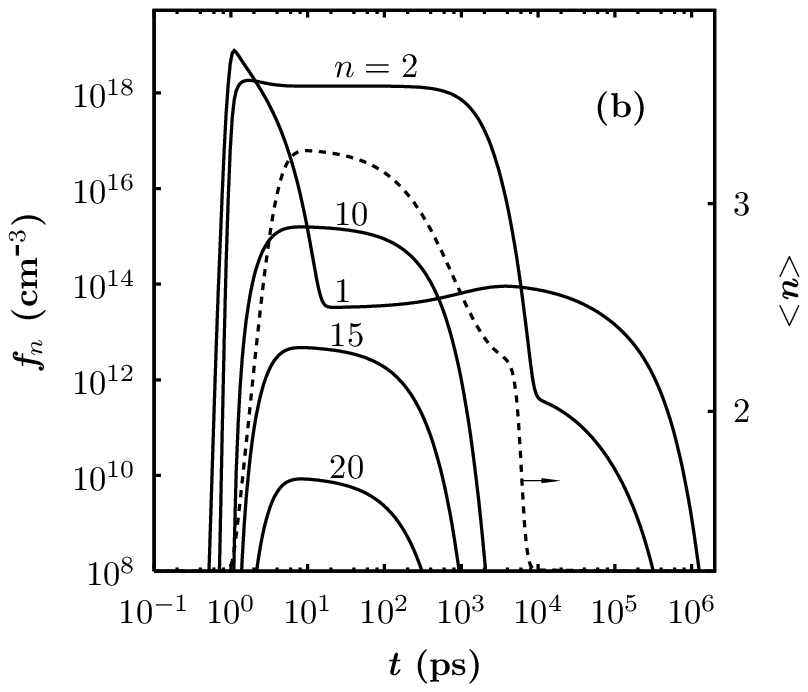,width=7.2cm,height=6cm,angle=0}}
  \caption{(a): Cluster concentration versus time and number of 
      e-h pairs per cluster for diamond at $T=12$ K under Gaussian
    pulse excitation  with $G_0 = 5.6\times 10^{31}$
    cm$^{-3}$s$^{-1}$, $t_p = 0.1$ ps, and $t_0 = 1.0$ ps. 
    (b): Concentration of some 
    selected clusters   (solid curve)
    and ANPC $\langle n\rangle$ (dashed curve) versus time.
    }
  \label{fig3}
\end{figure}

We now discuss   kinetics under a higher excitation.
Fig.\ 3 shows the same calculation as 
in Fig.\ 1 but with a much higher excitation, {\em i.e.}, ninety times as
large as in Fig.\ 1. This intensity corresponds to an excitation
of $\sim$17\ mJ/cm$^2$  in the experiment.\cite{PRB} 
It is seen from the figure that compared to the case of low excitation
in Fig.\ 1,
the exciton  density  decays  during  the first 20\ ps.
This explains the absence of  exciton luminescence in the
experiment performed at a similar excitation.\cite{PRB}
It is also seen that although the peak exciton density is 
almost proportional to the excitation intensity, the exciton density in
the quasi-equilibrium is smaller for larger excitation
intensities.
In Fig.\ 1(b) and Fig.\ 3(b), the exciton density in
quasi-equilibrium is about $10^{15}$\ cm$^{-3}$ under an 
excitation of 0.2\ mJ/cm$^2$ 
and only  $10^{13}$ cm$^{-3}$ under 17 mJ/cm$^2$. 
Moreover the ANPC increases very little despite such a large 
increase in excitation intensity. This behavior confirms the second-order nature 
of the exciton-EHD transition at this low temperature, and it
is similar to  what was discovered in Ge and Si at sufficiently low 
temperatures.\cite{2nd_order2} 
It is understood that the larger
excitation intensity makes the concentration of  clusters
grow up more rapidly  since the gain  $g_n$ 
in Eq.\ (\ref{gn}) is proportional to the exciton density. 

\begin{figure}
  \centerline{\psfig{figure=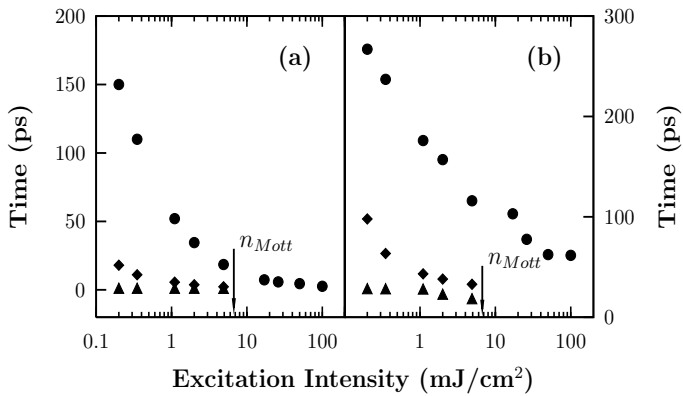,width=9.cm,height=5.2cm,angle=0}}
  \caption{Calculated (a) and experimental (b) e-h droplet formation
    times and exciton formation times versus the excitation intensity. 
    {\Large $\bullet$}: time used when $n_d$ reaches its maximum;
    {\small $\blacklozenge$}: time used when $n_d$ reaches its half maximum;
    {\normalsize $\blacktriangle$}: time used when $n_x$ reaches its maximum.
  }
  \label{fig4}
\end{figure}

Before we discuss the high temperature case, 
we analyze the dependence  on the excitation intensity at low
temperature. In Fig.\ 4 we compare 
the time delays at which $n_d$ reaches the maximum and half
maximum as well as  the time
delay needed for the exciton density $n_x$
to reach a maximum under different 
excitation intensities [Fig.\ 4(a)] with those in the experiment
[Fig.\ 4(b)].\cite{PRB} 
The figure shows that our results are in good qualitative agreement
with experiment.
In both figures the formation time of e-h cluster
decreases with the increase of the excitation intensity, and the time when 
the exciton density (luminescence intensity) reaches its maximum is 
independent of the excitation intensity. 
One may find that  
the time of the maximum density of exciton in our results is nearly 
zero which is smaller than the experimental value around 20
ps. This is due to the fact that 
we use a simple model of Gaussian excitation in
which the relaxation process of the e-h system is neglected. 
As mentioned
before, the relaxation, diffusion, and the Auger processes
slow down the EHD formation process. Thus our calculations produce a relatively
small formation time.

\begin{figure}[htb]
  \centerline{\psfig{figure=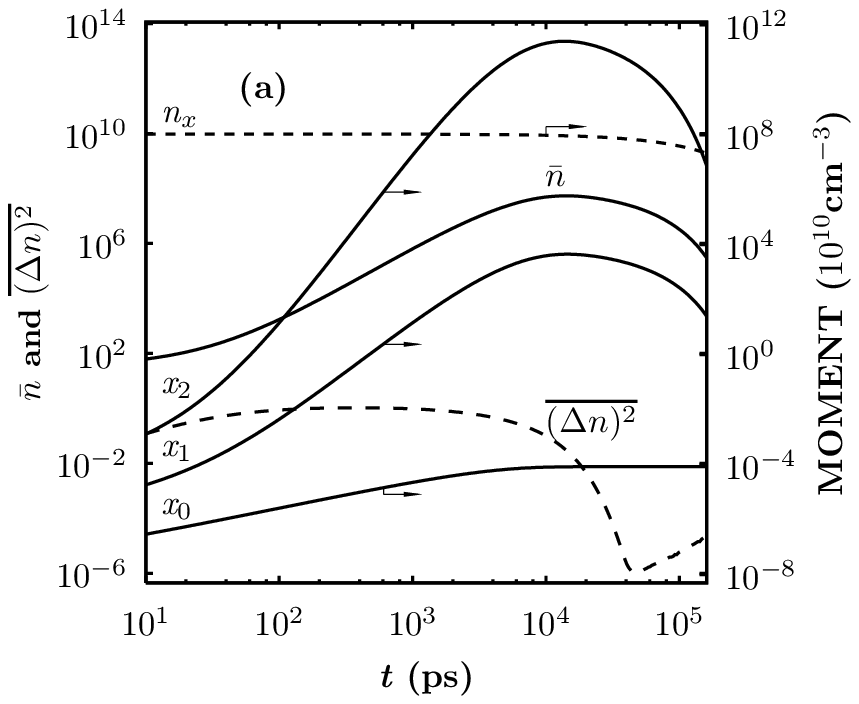,width=7.2cm,height=6.cm,angle=0}}
  \bigskip
  \centerline{\psfig{figure=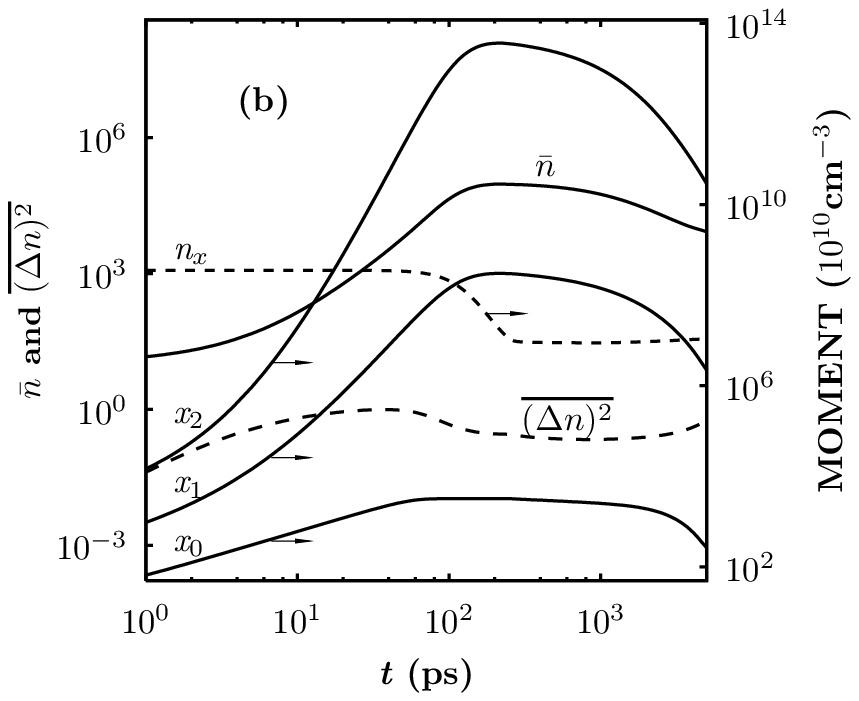,width=7.2cm,height=6.cm,angle=0}}
  \caption{Time dependence of the ``exciton density'' $n_x$, the
    moments of EHD distribution, $x_0$,$x_1$,$x_2$, for diamond at 100
    K, the average number of e-h pairs per drop, $\bar{n}$ and the
    relative  mean square of the droplet distribution 
    $\overline{(\Delta n)^2}$ under different excitation. We use
    Gaussian pulse excitation  with $t_p = 0.1$ ps and $t_0 = 0.1$ ps:
    (a): $G_0 = 6\times 10^{30}$ cm$^{-3}$s$^{-1}$; 
    (b): $G_0 = 20\times 10^{30}$ cm$^{-3}$s$^{-1}$.
  } 
  \label{fig5}
\end{figure}

We now study the kinetics of diamond at high
temperature (100\ K) using equations of moment Eqs.\ (\ref{12}-\ref{cont}). 
It is noted that this method only allows one to study the formation
and decay processes close to 
the steady state. We 
follow the path used for the  
low temperature case, {\em i.e.}, first a small excitation and
then  a high excitation.

In Fig.\ 5(a) we present the evolution  of the
 ``exciton density'' Eq.\ (\ref{nxt}), 
the three integer moments of the EHD distribution Eq.\ (\ref{moment}), 
the average size of EHD $\bar n=x_1/x_0$ and the
relative mean square of EHD distribution 
\begin{equation}
  \overline{(\Delta n)^2} = [x_2/x_0-(x_1/x_0)^2]/(x_1/x_0)^2
\end{equation}
under an excitation of 1.8\ mJ/cm$^{2}$.
Very different from the results at low temperature where only
small clusters are formed, here one finds that the average drop
size is very large: about $10^7$. 
Nevertheless the density of all e-h pairs which are condensed 
in EHDs, {\em i.e.} $x_1$, is rather small,
less than $10^{13}$\ cm$^{-3}$.
This can be understood easily: 
For a high temperature the evaporation rate $\alpha_n$ is much larger
and this larger evaporation 
impedes  the formation of EHD.
For the same reason the formation of EHD slows down.
The time when EHD density reaches its maximum is about $10^4$\ ps 
compared to about 40\ ps under the same excitation at 12\ K. The
relative mean square of EHD distribution $\overline{(\Delta n)^2}$,
which describes the fluctuation of droplet distribution, is very small
when the system is in the steady state. 

The kinetics at a  higher excitation intensity of 6.1\
mJ/cm$^{2}$ is plotted in Fig.\ 5(b).
Note that the curve of $x_1$  in Fig.\ 5(b) is comparable with
that in Fig.\ 4(b) of  
Ref.\ \onlinecite{High Tc}---where the excitation 
density and temperature are similar to
those in the calculation---except that
our result is plotted in a logarithmic scale and in a larger time range.
Moreover the kinetics shows two main differences from the low
excitation  case: 
First, the average drop size $\bar{n}$ is much smaller: about $10^4$; 
Second, the density of all e-h pairs which are condensed into
droplets, $x_1$, is much larger---around $10^{18}$\ cm$^{-3}$ which is
at least four orders of magnitude larger. 

The high excitation creates a large number of excitons, which 
 then produce a large number of small
droplets. However this process, together with the growth of the newly
created small droplets, tend to reduce $n_x$. 
As $n_x$ becomes smaller, the thermal potential in Eq.\ (A1) 
forms a higher barrier between  excitons/multiexciton complexes  and
EHDs.\cite{WestP} When the thermal potential grows  high enough, it
blocks the multiexciton complexes to grow into small droplets and the reverse 
process. Thus for a relatively long time, the droplet density $x_0$
becomes stable. It is noted that $x_0$ always becomes stable before other 
moments, $x_1$ and $x_2$, get stable as shown in 
Figs.\ 5(a) and (b).\cite{Haug3 indirect-gap1,Staehli} 
Thus the formation process is separated into
two stages:  The first is the process of the growth of the number of
droplets; and the second is the process of the growth 
of the size of the droplets. From Fig.\ 5(a)
one can see that 
the first stage ends at about $8\times 10^3$ ps, and the second one
ends at $2\times 10^4$ ps. The formation process after the  first stage
is a key one for the growth of the size of droplets. 
In Fig.\ 5(a) one can see
that $n_x$ stays nearly  
unchanged during this process, because the number of droplets is very
small and the growth of these droplets requires very few
excitons. In this 
case the thermal potential in Eq.\ (A1) remains nearly the same 
so that there is enough time for the e-h system to evolve
slowly into its equilibrium,  
where  the average drop size is very large,
while in Fig.\ 5(b) one finds
 the process in 80-200\ ps causes $n_x$ to decrease by 
one order of magnitude.
The depletion of excitons prevents droplets to grow larger, 
{\em i.e.}, the shortage of
excitons stops the growth of the size of droplets when the 
gain  $g_n$ and loss $l_n$ rates are equal. 
Therefore the system can only reach a steady state which
is in fact far away from equilibrium.
It is  noted from Eq.\  (\ref{n_c}) that $n_c \propto (\ln
n_x)^{-3}$, {\em i.e.}, $n_c$ decreases with the increase of 
excitation.
This, together with the fact that the  formation rates 
of multiexciton complexes are 
proportional to $n_x$, tend to increase the number of small droplets
greatly with the increase of excitation density.
In short, a large excitation tends to create too many small droplets 
which are unable to grow into large ones
due to the limited number of  excitons. And as a
result, the density of exciton, $n_x$, is small while the density 
of all e-h pairs condensed in droplets, $x_1$, is large.
These metastable features are similar to those in Ge at 
a high enough temperature.\cite{Staehli} 

In summary we have studied the kinetics of EHD formation and decay at
low (12 K) and high (100 K)
temperatures under both low and high excitations by master equations.
At low temperature our results are comparable with  measurements
reported by Nagai {\em et al.}\cite{PRB} The time evolution of 
exciton and EHD basically represents the time-resolved photo-luminescence 
measurement. The possible
causes of the discrepancies between the calculation and the experiments
are discussed in detail.
The ANPC under both low and high excitation are too
small to form  macroscopic EHDs. The
smallness of ANPC and the  time 
during which the system is in
equilibrium indicate that the phase 
transition is a second order process as  in direct
semiconductors. 
Despite much simplification in the model of the master equation
theory, our results are in good
qualitative agreement with  experimental results. 
Our study of EHD at high temperature predicts that the average drop size 
is as large as $10^6$. Moreover, under low excitation
the average size of EHDs is very large 
 but the EHD density is very low. Under high excitation 
the average size of EHDs is much smaller, but the EHD density
is much larger. The physics behind these predictions is
discussed in detail.
These effects demonstrate the metastable feature of the
kinetics at high temperature in diamond.
Experiments are needed to verify these predictions.

This work was supported by the Natural Science Foundation of China 
under Grant No. 90303012. 
MWW was also supported by the ``100 Person Project''
of Chinese Academy of Sciences and the Natural Science Foundation of China 
under Grant No. 10247002.
MKG acknowledges JSPS, KAKENHI (S) and SORST program from JST  for financial 
support. The authors gratefully acknowledge the critical reading 
of this manuscript by Dr. Jean Benoit Heroux.
JHJ would like to thank L. Jiang for valuable discussions.

\begin{appendix}

\section{The expressions of $J_{n_c}$ and $f(n_c,t)$}
The probability current from droplets $J_{n_c}$ 
in Eq.\ (8) is given by Staehli:\cite{Staehli}
\begin{equation}
  J_{n_c} \approx b n_x n_c^{2/3} [ \frac{F_x(n_c)}{p} -
  (\frac{n_c}{\bar{n}})^{3/2} \frac{C}{p}
  \exp(\frac{\Psi(\bar{n})-\Psi(n_c)}{kT})]\ ,
\end{equation}
with $F_x(n)=n_x n^{3/2} \exp[-\Psi(n)/kT]$\ denoting the
distribution function of excitons and multiexciton complexes. 
$p = \sqrt{2\pi kT/(\partial ^2 \Psi(n)/\partial n^2)_{n_c}}$\ is the
width of the potential barrier between excitons/multiexciton complexes and EHDs.
$\bar{n} = x_1/x_0$\ is the average drop size. $C=x_0/[2
\int\limits^{\bar{n}}_{n_c} (n/\bar{n})^{3/2}\exp ( \frac{\Psi(\bar{n})
-\Psi(n)}{kT})\,dn]$\ is a normalization factor of the distribution
function of EHDs and the thermal potential $\Psi(n)$\ is given
by\cite{WestP}
\begin{equation}
  \Psi(n) = -kT n \ln\frac{n_x}{n_s} + 4\pi R_0^2 \sigma n^{2/3}
  + kT \sum\limits^{n}_{j=1} \ln (1 + \frac{j}{\alpha _j \tau
  _d})\ .
\end{equation} 

As for $f(n_c,t)$, when $\frac{\partial n_c}{\partial t} > 0$ 
it takes the form of distribution function of excitons and
multiexciton complexes: 
\begin{equation}
  f(n_c,t) \approx F_x(n_c)/2\ .
\end{equation}
Otherwise it takes the form of distribution function of EHDs:
\begin{equation}
  f(n_c,t) \approx C (\frac{n_c}{\bar{n}})^{3/2} \exp
         (\frac{\Psi(\bar{n})-\Psi(n_c)}{kT})\ .
\end{equation}

\end{appendix}


\begin{thebibliography}{10}
\bibitem{60K} The boundary temperature $T_b$ can also be evaluated 
  by scaling:  For Si, $T_b$ is approximate 9 K and the work function
  $\phi=95.1$\ K. For diamond, $\phi=580$\ K and the recombination
  rate is 150 times as large as that of Si.
  Therefore from
  $T_b^2\exp(-\phi/kT_b)|_{\mbox{diamond}} = 
  150\times T_b^2\exp(-\phi/k
  T_b)|_{\mbox{Si}}$, we conclude $T_b|_{\mbox{diamond}} \approx 60$ K.
\bibitem{exciton} We assume 100 ns for exciton lifetime in our 
  calculation. Exciton lifetime depends on the 
  temperature, excitation density, and sample quality---because lifetime
  is mainly determined by impurity-related
non-radiative processes. 
  However the simulation including these effects is too complicated. 
  We use the approximation of 100 ns, which does not make essential 
  difference in the time region of about 100 ns.
\bibitem{parameter} This is the work function at 100 K. 
  The work function at 12 K may be smaller as the experiment by
  Sauer {\em et al.}\cite{Thonke2} suggested. However, even if one 
  takes the work function at 12 K as one half of that at 100 K 
  to evaluate evaporation rate at 12 K, the calculation shows that
  the evaporation rate at 12 K is so small that one can drop the
  evaporation term in the calculation.
\bibitem{exci} We assume that the excitation density depends linearly
  on exciting laser intensity with $G_0 = I_x \ \times \ 0.33\times 
  10^{31}$ cm$^{-3}$s$^{-1}$, where $I_x$ is excitation intensity in
mJ/cm$^2$(The coefficient $0.33\times 10^{31}$ cm$^{-3}$s$^{-1}$ 
  is obtained by fitting $n_{Mott}$ to exciting laser intensity 
  $6.9$ mJ/cm$^2$ in Fig.\ 4. of Ref.\ \onlinecite{PRB}).
\end{thebibliography}
\end{document}